\documentclass[12pt,preprint]{aastex}

\def\tighttable{\def\baselinestretch{1.0}}
\def\arcsec{\ifmmode '' \else $''$\fi}
\def\arcmin{\ifmmode ' \else $'$\fi}
\def\arcsecpoint{\ifmmode ''\!. \else $''\!.$\fi}
\def\arcminpoint{\ifmmode '\!. \else $'\!.$\fi}
\def\cc{\ifmmode {\rm cm}^{-3} \else cm$^{-3}$\fi}
\def\cl{\ifmmode {\rm cm}^{-2} \else cm$^{-2}$\fi}
\def\micron{\ifmmode \mu{\rm m} \else $\mu$m\fi}
\def\kms{\ifmmode {\rm km\,s}^{-1} \else km\,s$^{-1}$\fi}
\def\Hubble{\ifmmode {\rm km\,s}^{-1}\,{\rm Mpc}^{-1}
	\else km\,s$^{-1}$\,Mpc$^{-1}$\fi}
\def\ergsec{\ifmmode {\rm ergs\;s}^{-1} \else ergs s$^{-1}$\fi}
\def\ergscm{\ifmmode {\rm ergs\,s}^{-1}\,{\rm cm}^{-2}
	  \else ergs\,s$^{-1}$\,cm$^{-2}$\fi}
\def\ergscmA{\ifmmode {\rm ergs\,s}^{-1}\,{\rm cm}^{-2}\,{\rm \AA}^{-1}
	  \else ergs\,s$^{-1}$\,cm$^{-2}$\,\AA$^{-1}$\fi}
\def\ergscmHz{\ifmmode {\rm ergs\,s}^{-1}\,{\rm cm}^{-2}\,{\rm Hz}^{-1}
	  \else ergs\,s$^{-1}$\,cm$^{-2}$\,Hz$^{-1}$\fi}
\def\Msun{\ifmmode M_{\odot} \else $M_{\odot}$\fi}
\def\qo{\ifmmode q_{0} \else $q_{0}$\fi}
\def\Ho{\ifmmode H_{0} \else $H_{0}$\fi}
\def\ltsim{\raisebox{-.5 ex}{$\;\stackrel{<}{\sim}\;$}}
\def\gtsim{\raisebox{-.5 ex}{$\;\stackrel{>}{\sim}\;$}}

\newcommand {\lya}{Ly$\alpha$}

\shorttitle{Galaxies around z=5.8 quasar}
\shortauthors{Zheng et al.}
\slugcomment{To appear in {\it The Astrophysical Journal}}
\received{2005 July 11}
\begin{document}

\title{An Overdensity of Galaxies near the Most Distant \newline 
Radio-Loud Quasar\altaffilmark{1}} 

\author{
W. Zheng\altaffilmark{2},
R. Overzier\altaffilmark{3},
R. J. Bouwens\altaffilmark{4},
R. L. White\altaffilmark{5},
H. C. Ford\altaffilmark{2},
N. Ben\'{\i}tez\altaffilmark{6},
J. P. Blakeslee\altaffilmark{2},
L. D. Bradley\altaffilmark{2},
M. K. Jee\altaffilmark{2},
A. R. Martel\altaffilmark{2},
S. Mei\altaffilmark{2},
A. W. Zirm\altaffilmark{3},
G. D. Illingworth\altaffilmark{4},
M. Clampin\altaffilmark{7},
G. F. Hartig\altaffilmark{5},
D. R. Ardila\altaffilmark{8},
F. Bartko\altaffilmark{9}, 
T. J. Broadhurst\altaffilmark{10},
R. A. Brown\altaffilmark{5},
C. J. Burrows\altaffilmark{11},
E. S. Cheng\altaffilmark{12},
N. J. G. Cross\altaffilmark{13},
R. Demarco\altaffilmark{2},
P. D. Feldman\altaffilmark{2},
M. Franx\altaffilmark{3},
D. A. Golimowski\altaffilmark{2},
T. Goto\altaffilmark{14},
C. Gronwall\altaffilmark{15},
B. Holden\altaffilmark{4},
N. Homeier\altaffilmark{2},
L. Infante\altaffilmark{16},
R. A. Kimble\altaffilmark{6},
J. E. Krist\altaffilmark{17},
M. P. Lesser\altaffilmark{18},
F. Menanteau\altaffilmark{2},
G. R. Meurer\altaffilmark{2},
G. K. Miley\altaffilmark{3},
V. Motta\altaffilmark{2,16},
M. Postman\altaffilmark{5},
P. Rosati\altaffilmark{19}, 
M. Sirianni\altaffilmark{5}, 
W. B. Sparks\altaffilmark{5}, 
H. D. Tran\altaffilmark{20}, 
and 
Z. I. Tsvetanov\altaffilmark{2}
}

\altaffiltext{1}{Based on observations with the NASA/ESA Hubble Space Telescope, 
obtained at the
Space Telescope Science Institute, which is operated by the Association of 
Universities of Research in Astronomy, Inc., under NASA contract NAS5-26555}
\altaffiltext{2}{Department of Physics and Astronomy, Johns Hopkins
University, 3400 North Charles Street, Baltimore, MD 21218}
\altaffiltext{3}{Leiden Observatory, Postbus 9513, 2300 RA Leiden,
Netherlands}
\altaffiltext{4}{UCO/Lick Observatory, University of California, Santa
Cruz, CA 95064}
\altaffiltext{5}{STScI, 3700 San Martin Drive, Baltimore, MD 21218}
\altaffiltext{6}{Inst. Astrof\'{\i}sica de Andaluc\'{\i}a (CSIC), 
  Camino Bajo de Hu\'etor, 24, Granada 18008, Spain}
\altaffiltext{7}{NASA Goddard Space Flight Center, Code 681, Greenbelt, MD 20771}
\altaffiltext{8}{Spitzer Science Center, IPAC, MS 220-6, California Institute of
Technology, Pasadena, CA 91125}
\altaffiltext{9}{Bartko Science \& Technology, 14520 Akron Street, 
Brighton, CO 80602}	
\altaffiltext{10}{Racah Institute of Physics, The Hebrew University,
Jerusalem, Israel 91904}
\altaffiltext{11}{Metajiva, 12320 Scenic Drive, Edmonds, WA 98026}
\altaffiltext{12}{Conceptual Analytics, LLC, 8209 Woburn Abbey Road, Glenn Dale, MD 20769}
\altaffiltext{13}{Royal Observatory Edinburgh, Blackford Hill, Edinburgh, EH9 3HJ, UK}
\altaffiltext{14}{Institute of Space and Astronautical Science, 3-1-1 Yoshinodai, 
Sagamihara, Kanagawa 229-8510, Japan}
\altaffiltext{15}{Department of Astronomy and Astrophysics, The
Pennsylvania State University, 525 Davey Lab, University Park, PA
16802}
\altaffiltext{16}{Departmento de Astronom\'{\i}a y Astrof\'{\i}sica,
Pontificia Universidad Cat\'{\o}lica de Chile, Casilla 306, Santiago
22, Chile}
\altaffiltext{17}{Jet Propulsion Laboratory, M/S 183-900, 4800 Oak Grove Drive, Pasadena, CA 91109}
\altaffiltext{18}{Steward Observatory, University of Arizona, Tucson,
AZ 85721}
\altaffiltext{19}{European Southern Observatory,
Karl-Schwarzschild-Strasse 2, D-85748 Garching, Germany}
\altaffiltext{20}{W. M. Keck Observatory, 65-1120 Mamalahoa Hwy., 
Kamuela, HI 96743}

\begin{abstract}
A five square arcminute region around the luminous radio-loud quasar SDSS 
J0836+0054 ($z=5.8$) hosts 
a wealth of associated galaxies, characterized by very red 
($1.3 < i_{775} - z_{850} < 2.0$) color. 
The surface density of these $z\sim5.8$ candidates is approximately 
six times higher than the number expected from deep ACS fields.
This is one of the highest galaxy overdensities at high redshifts, which may 
develop into a group or cluster. 
We also find evidence for a substructure associated with one of the candidates.
It has two very faint companion objects within two arcseconds, which are 
likely to merge.
The finding supports the results of a recent simulation that luminous quasars 
at high redshifts lie on the most prominent dark-matter filaments and are 
surrounded by many fainter galaxies. The quasar activity from these regions 
may signal the buildup of a massive system. 
\end{abstract}


\section{INTRODUCTION}

The fluctuations in the cosmic microwave background temperature observed by 
{\it WMAP} \citep{map} are believed to be the seeds for the first generation 
of baryonic objects, which eventually evolved into the stars, galaxies and 
clusters seen in the current epoch. 
Several hundreds galaxy candidates at $z \gtsim 6$ have been found as 
i-dropout objects \citep{stanway,yan,bouwens,bouwens3}, 
characterized by a large color difference between the $i_{775}$- and 
$z_{850}$-band and faint magnitudes of approximately 26.
The current best estimate 
of the average surface density of these galaxies is from the
GOODS survey \citep{goods2}: approximately 0.25 galaxy per 
square arcminute to a $z_{850}$-band magnitude of 26.5
\citep{goods,bouwens3}. 
In addition, at least a dozen of 
quasars have also been discovered at redshift 
$z \gtsim 5.7$ \citep{fan2}, and their high luminosities suggest the presence 
of massive black holes of $> 10^9$ \Msun. 
The formation of such black holes at epochs less than one billion years after 
the Big Bang requires an extremely high accretion rate \citep{haiman}. 

According to the hierarchical model of galaxy formation \citep{cdm}, 
massive baryonic objects are formed in the densest halos. 
The latest simulations of CDM growth \citep{lcdm}
predict that $z\sim 6$ quasars may lie in the center of very massive 
dark matter halos of $\sim 4 \times 10^{12}$ \Msun, which will grow into
the most massive clusters of $\sim 4 \times 10^{15}$ \Msun\ at $z=0$. 
Studying the clustering properties of young galaxies and quasars 
may provide insights into their formation history. 
Reports of clustering of galaxies at redshift $z\sim 6$ 
\citep{ouchi,jxw,sangeeta} suggest that the 
formation of large structures in the early Universe may be more significant
than the cold dark matter (CDM) models have predicted. 
Recently \citet{massimo} reported that one of the most distant quasars, SDSS 
J1030+0524 at $z=6.28$, exhibits an excess of associated galaxies. It has 
also been known that some radio galaxies harbor an enhanced number of 
associated galaxies. \citet{venem}, and \citet{miley} 
found a potential protocluster around TN1338-1942,
a radio galaxy at $z=4.1$. \citet{venemans} and \citet{overzier} observed 
an excess of \lya\ emitters (LAEs) and Lyman break galaxies (LBGs) in the 
field of TN0924-2201, the most distant radio galaxy at $z=5.2$. These 
reports confirm an empirical result 
that powerful, high-redshift radio galaxies are associated with massive 
forming galaxies \citep{deb}. To explore the possibility that 
radio-loud quasars may also be a signpost of galaxy clustering at high 
redshifts, we initiated a project with HST/ACS to image the fields around 
some of the most distant radio-loud quasars.

The quasar SDSS J0836+0054 \citep[$z=5.82$;][]{fan} is the most distant 
radio-loud quasar known to-date, and is one of the most luminous. 
The object is detected in the FIRST radio survey,
with a total flux density at 1.4 GHz of 1.11 mJy.
VLA and MPMBA-IRAM observations \citep{petric} yield  flux densities of 
1.75 mJy at 1.4 GHz, 0.58 mJy at 5 GHz, and a non-detection at 250 GHz, with 
a $3\sigma$ upper limit of 2.9 mJy.
It is a compact steep-spectrum radio source with a radio spectral index of 
$-0.8$. VLBI observations at $\sim 10$ milliarcsecond angular resolution 
\citep{frey} show an apparent core-jet morphology, with no indication of 
multiple images produced by gravitational lensing.
The quasar's enormous power at $M\sim -27.8$ in the rest-frame
UV band implies a mass 
of the central black hole of $5 \times 10^9$ \Msun, 
posing a challenge to the theoretical models on how such massive objects are
formed in the very early Universe. 

Throughout the paper, we use AB magnitudes and assume the common values of 
cosmological parameters: ($\Omega_m,\Omega_\lambda$) = 
($0.3,0.7$) and \Ho\ = 70 \Hubble.

\section{DATA}

The HST/ACS observations of SDSS J0836+0054 were carried out on 2004 October 8 
and November 17, with a total exposure of 10778 sec in the $z_{850}$-band 
and 4676 sec in the $i_{775}$-band, respectively, at 
two different position angles. 
They were processed with the standard pipeline CALACS \citep{calacs}, 
then with APSIS \citep{apsis}. 
These procedures carried out flat-fielding, removing bias, dark-current and 
cosmic-ray events, corrections for geometrical distortion of the 
detectors, and drizzling the dithered images. The final images  
cover approximately 11.4 square arcminutes. The APSIS tasks determine 
the Galactic extinction of  E(B-V)=0.05 from the dust maps of \citet{dust}
and applied 0.1 and 0.07 to $i_{775}$- and $z_{850}$-band magnitudes, 
respectively. 

We used SExtractor \citep{sex} to find and parameterize sources
from the science images and their RMS counterparts. We first used the 
$z_{850}$-band as the detection image and then reran the task in a dual 
mode, namely to use the profile information in the $z_{850}$-band to link and 
constrain the parameter counterparts in the $i_{775}$-band image. 
The limiting magnitudes are similar to those
of the GOODS fields: 26.5 in the $z_{850}$-band, for a 
$10\sigma$ detection of a source of 0\farcs 2 in diameter.
As shown in Table 1, we selected objects with a color 
$i_{775} - z_{850} > 1.3$ {\sf (MAG\_ISO)}. 
Only sources with a star-galaxy index of $<0.8$ were considered 
(0 for galaxy and 1 for star). To further avoid contaminations from cosmic 
ray events, we only considered sources in the sky region that is covered 
by all six exposures in each band, which is approximately 10 square 
arcminutes.
The limiting $z_{850}$ magnitudes are {\sf MAG\_AUTO} as they allow us to 
collect most
of the source flux, but the colors are determined with {\sf MAG\_ISO}, 
in order to maximize the signal-to-noise (S/N) ratio. Extensive tests by 
\citet{bpz04} suggest that the colors of 
faint galaxies are more robust with {\sf MAG\_ISO}.
The color selection threshold of 1.3 is chosen to avoid the 
contamination from low-redshift interlopers. 
Seven sources are listed, and five of them are $i_{775}$-faint (i-faint) 
objects, enabling us 
to separate them from objects at $z>6$. \lya\ emission at $6<z<7$ is at 
wavelengths redward of the $i_{775}$-band, and the $i_{775}-z_{850}$ color is 
at least
2.5 magnitudes.  No detection in the $i_{775}$-band is therefore anticipated.
There are nine objects with $1.0 < i_{775}-z_{850} < 1.3$, which may be LAEs 
at $z\sim 5.8$, or galaxies at $z\sim 1$.

At a redshift of 5.8, one arcsecond corresponds to approximately 5.7 kpc,
and the transverse dimension of the ACS field is approximately 1.1 Mpc. 
Along the line of sight, the proper distance between $z=5.7$ and 5.9 
is approximately 13 Mpc. 

\section{RESULTS}

We identified seven galaxy candidates with $z>5.5$ 
in the field of this quasar. 
Table 1 lists the properties of the candidates, along with that of 
the quasar, and Fig. 1 displays their positions. Object D is 
near a foreground cluster approximately one arcminute south of the quasar. 
The cutout images of each candidate are shown in Fig. 2. 
We carried out a test on negative images \citep{goods} as spurious sources 
may be present even at S/N $> 5$. The $i_{775}$- and $z_{850}$-band images were 
multiplied by $-1$, and SExtractor tasks were carried out using the same detection 
parameters. No candidate was detected with the same selection criteria.

The Bayesian photometric redshifts \citep[BPZ;][]{bpz} were calculated with 
a calibrated template set of \citet{bpz04} supplemented by a very blue 
starburst template. These templates significantly improve the photometric 
redshift estimation for faint, high-$z$ galaxies. 
The new features in BPZ identify multiple peaks of redshifts 
and assigns a probability to each of them. The values of the first-peak
redshifts are listed in column 8 of Table 1.

While it is common to refer to objects with large $i_{775}-z_{850}$ color as 
``i-dropouts'', we can obtain a better redshift discrimination by selecting 
objects with strong $i_{775}-z_{850}$ breaks, but which are still detected in 
the $i_{775}$-band. At $z\sim 5.8$, the redshifted \lya\ feature straddles both
the $i_{775}$- and $z_{850}$-band. 
Because of the high throughput of ACS, most candidates at this redshift are 
expected to be detected in the $i_{775}$-band. As the redshift increases from 
5.8 to 6.0, \lya\ emission rapidly moves out of the $i_{775}$-band, and 
therefore the $i_{775}-z_{850}$ 
color increases rapidly from magnitude $\sim 1.5$ to $\gtsim 2.0$. 
Having set a detection limit of $z_{850} < 26.5$, we are able to secure a 
$2\sigma$ detection of sources with $i_{775}-z_{850} < 2.0$. 
We tested the reality of $i_{775}$-band detections by deliberately shifting the
$i_{775}$-band image by 
$\pm20$ pixels along its row and column and then reran SExtractor.
The source apertures defined by the $z_{850}$-band images would point to a
nearby sky field, and in all the cases, the aperture source fluxes in the 
$i_{775}$-band 
were below S/N=1.4. We therefore conclude that an $i_{775}$-band detection at 
S/N $>2$ 
is unlikely the result of random fluctuations. The detection in both bands 
significantly enhances the confidence level, and also enables us to exclude 
background galaxies at $z>6$ ($i_{775}$- or i-dropouts). 
Objects B and G in Table 1 are not 
detected in the $i_{775}$ band. They are likely background galaxies at $z>6$ 
and not physically associated with the quasar.

\section{DISCUSSION}

The GOODS results \citep{goods,bouwens3} suggest that the surface density of 
i-dropouts and i-faint objects is approximately 0.25 per square arcminute, to 
a limiting magnitude of $z_{850}\sim 26.5$. These objects are believed to be 
at redshifts $5.5 < z < 6.5$. Since we only select i-faint objects, the
photometric redshifts of these objects only cover the range of $5.5< z < 6$.
From previous studies \citep{goods,massimo,bouwens3} we 
estimate that approximately 60\% of these objects are i-faint and the rest 
i-dropouts. The surface density of i-faint objects is only $\sim 0.15$ per 
square arcminute, or 1-2 in the ACS field of view. 
Since the chance of finding seven i-dropouts (including i-faint objects) in 
GOODS in a random ACS/WFC-size cell ($\sim11$ square arcminutes) amounts to a 
few percent \citep{massimo}, our finding of five i-faint objects (and 
two i-dropouts) in one ACS field suggests a significant source overdensity. 
Since all our candidates lie in a region of five square arcminutes the actual 
factor of overdensity is approximately six with respect to GOODS, with no 
random cells drawn from GOODS containing seven objects \citep[four being the 
highest using the sample of][]{bouwens3}. Although cosmic variance is expected 
to be non-negligible even on scales as large as probed by GOODS \citep{var}, 
field-to-field variations can not be determined empirically until larger 
surveys become available.

Our results provide new evidence for an excess of galaxies associated with 
quasars at $z\gtsim 5.8$. \citet{massimo} find seven candidates at $z\gtsim 6$
in the field of 
a radio-quiet quasar at $z=6.28$. Four out of these seven candidates are 
detected in the $i_{775}$-band, and are likely to be foreground galaxies at 
redshift $5.5 < z <6$. The number of galaxies associated with the
$z=6.28 $ quasar is at most three. We therefore believe that the five $z\sim 
5.8$ candidates in the vicinity of quasar SDSS J0836+0054 represents a 
significant overdensity.  Such an overdensity is consistent with the prediction of 
the Millennium Simulation \citep{lcdm} that a 'first quasar' candidate at $z=6.2$ 
lies on one of the most prominent dark-matter
filaments and is surrounded by a large number of other, much fainter galaxies. 
The quasar itself exhibits an $i_{775}-z_{850}$ color of 1.2, which is considerably 
smaller than that in the SDSS (2.2). This is because the $i_{775}$-band sensitivity of   
ACS detectors extends further to the red than that of the SDSS. The bulk of 
redshifted \lya\ emission falls into that extended wavelength
region beyond 8000 \AA\ and boosts the $i_{775}$-band flux.
We carried out color simulations of galaxies at $z \sim 5.8$, using the 
spectral evolution 
library of \citet{model}. Among the 39 representative model templates, we 
choose 21 at age of 1.4 Gyr and younger, which include star-burst models. 
We also added \lya\ emission lines to the models with ages of $25$ Myr. 
The \lya\ emission 
is assumed to have a line width FWHM=1000 \kms, and varying equivalent 
widths 100 or 200 \AA\ in the rest frame. All the model spectra are 
redshifted, and corrected for intergalactic absorption. 
As shown in Fig. 3, the $i_{775}-z_{850}$ color peaks around 1.7 for 
LBGs, and that of LAEs extends at $1.0 < i_{775}-z_{850} < 1.6$.
Approximately 90\% of the LBGs are selected by our criteria, but only a 
fraction for LAEs.

The field around source C is interesting. The source seems to be a multiple 
system, as shown in Fig. 4. In its vicinity there are two red objects, which 
are slightly fainter and are marked as C2 and C3. 
The two companion objects of source C are not among our initial sample, 
as they do not meet our selection criteria. However, they are in the vicinity 
of source C and share similarly red colors suggesting that they may be 
physically associated. The angular separation between these three sources is 
only $\sim 1.7$ arcsecond or $\sim 10$ kpc, suggestive of merging. 

There is a slight chance that some of the candidates are reddened objects at 
lower redshifts. 
Red galaxies at $z\sim 1$ display a color of $0.9 \ltsim i_{775}-z_{850} 
\ltsim 1.1$ \citep{mei}.
Uncertainties in the $i_{775}-z_{850}$ color in our sample may introduce 
contaminations from red galaxies at $z\sim 1$.
While it is possible that large uncertainties in magnitudes may lead to 
contamination from 
these galaxies, such an effect should also apply to the ACS images of the 
GOODS fields because of the same instrument and comparable exposure times. 
As the GOODS fields exhibit a considerably lower surface density of i-faint 
objects than our field, the contamination level is low at $\sim 8\%$ 
\citep{bouwens}.
The so-called Extremely Red Objects (ERO) can display sharply rising fluxes 
toward longer wavelengths in several adjacent bands. The lack of deep infrared 
images does not allow us to exclude such objects, but since 
the sizes and morphologies of our candidates do not fit either
type, having neither pointlike or rather diffuse profiles, we conclude that
they are unlikely EROs.
\citet{sangeeta} suggest that 
$\sim 90\%$ of sources with $i_{775}-z_{850}>1.3$ 
and $z_{850}<27$ are spectroscopically confirmed at redshift $z>5.5$.
The actual contamination rate is even lower when star-like objects are 
excluded. 

Our finding complements a number of recent reports of overdensities of 
associated galaxies in the vicinities of radio galaxies  at $z>4$ 
(although not exclusively) and suggest a possibility 
of enhanced activities of clustering in the fields of radio sources at high 
redshifts. The galaxy candidates are more than a hundred times fainter than the
quasar itself, and they form a good example of a hierarchical evolution. The 
detection of radio emission signals an environment that is rich in dark matter,
thus harboring the formation of massive black holes.

\section{SUMMARY}

The region around the radio-loud quasar SDSS J0836+0054 is rich in i-faint 
objects, which are candidates for galaxies at $z\sim 5.8$. The surface density 
of these objects is 
approximately 4-6 times higher than that of the GOODS fields, yielding one 
of the highest overdensities at $z \sim 6$ known to-date. 
Our finding supports a hierarchical structure predicted by a recent simulation 
that luminous quasars are surrounded by fainter galaxies. 
Spectroscopic observations are needed to further confirm the association and 
enable us to estimate the volume density occupied by these 
sources. Future observations of this field in infrared bands will also provide 
information about the spectral energy distribution of these distant sources. 

The observations add fresh evidence that quasars and radio galaxies are good 
beacons for finding protoclusters of young galaxies at high redshifts.
Our measurements could provide the first constraint on the halo occupation
number for LBGs in one of the most massive haloes at high redshift, which will 
provide interesting comparison to numerical simulations.

\acknowledgments

ACS was developed under NASA contract NAS 5-32865, and this research 
has been supported by NASA grant NAG5-7697 and by an equipment grant from Sun 
Microsystems, Inc. The Space Telescope Science
Institute is operated by AURA Inc., under NASA contract NAS5-26555.
We are grateful to K.~Anderson, J.~McCann, S.~Busching, A.~Framarini, 
S.~Barkhouser, and T.~Allen for their invaluable contributions to the 
ACS project at JHU.  We thank the anonymous referee for constructive comments.

\clearpage

{
\begin{deluxetable}{cccccccc}
\rotate
\tablecaption{Objects with Large $i_{775}-z_{850}$ Color\label{tbl-1}}
\tablewidth{0pt}
\footnotesize
\tighttable
\tablehead{
\colhead{Object} & \colhead{$\alpha_{J2000}$} & \colhead{$\delta_{J2000}$} &
\colhead{$z_{850}$} & \colhead{S/N ($i_{775}$)\tablenotemark{a}} & 
\colhead{$i_{775}-z_{850}$\tablenotemark{a}} & \colhead{FWHM (\arcsec)} & \colhead{BPZ\tablenotemark{b}}
}
\startdata
A & 08 36 45.248 & 00 54 10.99 &$25.54 \pm 0.10$ & 3.2& $1.91\pm 0.36$ & 0.46 & $5.8^{+1.4}_{-0.2}$\\ 
B\tablenotemark{c} & 08 36 47.053 & 00 53 55.90 & $ 26.00\pm 0.17$ & 1.5& $2.40 \pm 0.97 $ & 0.39 & $ 5.9^{+1.0}_{-1.0}$\\ 
C & 08 36 50.099 & 00 55 31.16 & $ 26.24 \pm 0.15$ & 2.4&$1.92 \pm 0.60$ & 0.29 &$ 5.9^{+1.1}_{-0.5}$\\ 
C2 & 08 36 50.058 & 00 55 30.54 & $ 27.26 \pm 0.39 $ & 1.5&$ 2.42\pm 1.23$  & 0.29 & $ 5.9^{+1.1}_{-1.5}$\\ 
C3 & 08 36 50.010 & 00 55 30.27 & $ 26.82 \pm 0.18$ & 0.7&$3.41\pm 3.43$\tablenotemark{d}  & 0.20 &$ 7.0^{+0.0}_{-0.7}$ \\ 
D & 08 36 48.211 & 00 54 41.19 & $ 26.42\pm 0.14$ & 2.2&$1.84 \pm 0.49 $& 0.29 & $ 5.8^{+1.2}_{-0.7}$\\ 
E & 08 36 44.029 & 00 54 32.79 & $26.39 \pm 0.16$ &2.3& $1.61 \pm 0.42 $& 0.19 & $ 5.2^{+1.7}_{-0.7}$\\ 
F & 08 36 42.666  & 00 54 44.00 &$ 26.03\pm 0.17$&2.6 &$ 1.64 \pm 0.42$& 0.46 & $5.7^{+1.2}_{-0.7}$\\ 
G\tablenotemark{c} & 08 36 45.962 & 00 55 40.53 &$ 26.36\pm 0.25$ & 1.7&$1.91\pm 0.63$ & 0.43 & $5.8^{+1.2}_{-0.8}$\\  
Quasar & 08 36 43.871& 00 54 53.15& $18.85 \pm 0.02$& 22 &$1.19 \pm 0.03$ & 0.11 &$ 5.7^{+0.1}_{-0.1}$\\ 
\enddata
\tablenotetext{a}{Calculated using {\sf FLUX\_ISO} and {\sf MAG\_ISO}.}
\tablenotetext{b}{Bayesian photometric redshift, at a 67\% confidence level. 
First-peak redshifts are estimated with a preset upper limit of $z=7.0$.}
\tablenotetext{c}{i-dropout. Not considered as being associated with the 
quasar.}
\tablenotetext{d}{Calculated using {\sf MAG\_AUTO}, as {\sf MAG\_ISO} yields 
no detection.}
\end{deluxetable}
}

\clearpage

\centerline{\bf Figure Captions}
\bigskip

\figcaption{Composite HST/ACS image of the field of the quasar SDSS J0836+0054.
The field size is approximately 3 arcminutes ($\sim 1.1$ Mpc of proper 
distance). The quasar (Q) and the candidates are marked with open circles.
\label{fig1}}

\figcaption{Cutout $i_{775}$- and $z_{850}$-band images of candidates.
The image sizes are 2 arcseconds. The objects are named alphabetically (see 
Table 1) as marked at the upper-left corner of each panel of paired images.
The images of the quasar, at the lower-right panel, are marked as 
``Q''. The band name is marked at the lower-left corner of each image.
The contour at the center of each image marks the area where flux is collected
for {\sf MAG\_AUTO}. The source color was calculated from {\sf MAG\_ISO}, 
which were derived from a compact core within the contour.
\label{fig2}}

\figcaption{Distribution of $i_{775}-z_{850}$ color for simulated galaxy 
spectra at redshift $z=5.8$. The open region represents LAEs, and the shaded
region LBGs. Our selection of $1.3 < i_{775}-z_{850} < 2.0$ 
is expected to include most LBGs, but only about 1/3 of the LAEs.
\label{fig3}}

\figcaption{Field around object C in the $z_{850}$ band. 
Two additional red objects are marked. The angular scale of one arcsecond 
($\sim 5.7 $ kpc of proper distance) is marked.
\label{fig4}}

\clearpage

\setcounter{figure}{0}
\begin{figure}
\plotone{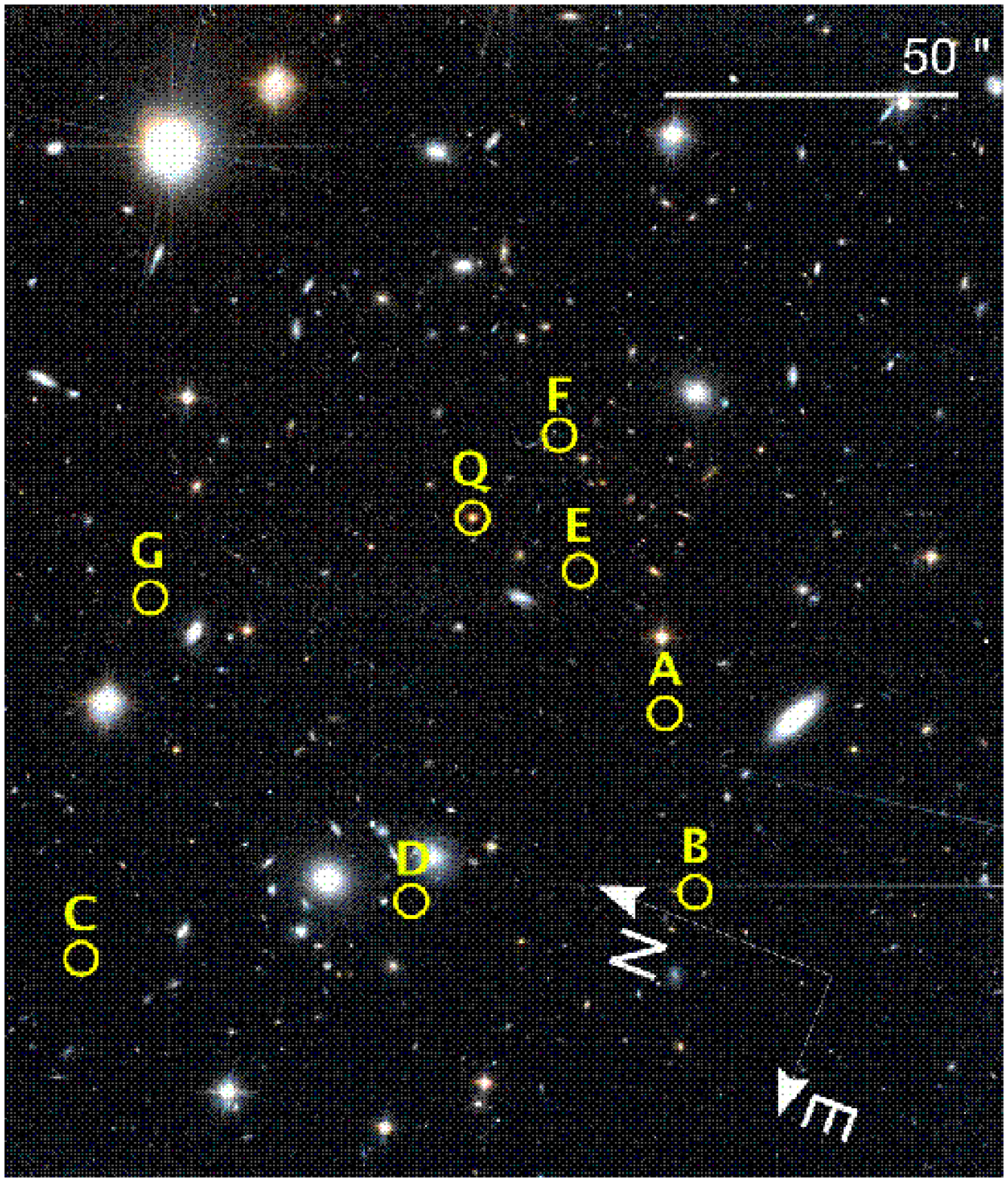}
\end{figure}

\begin{figure}
\plotone{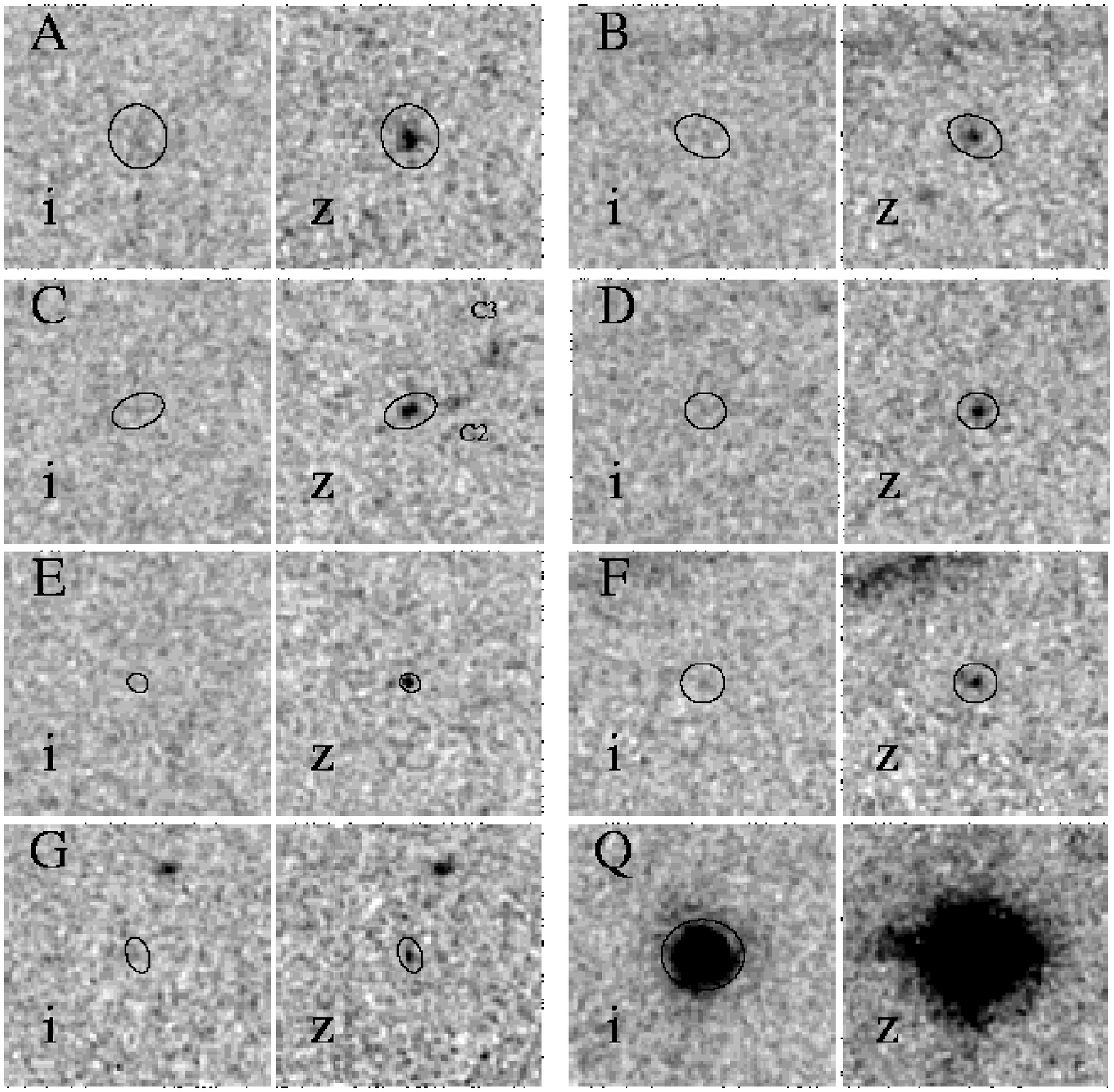}
\end{figure}

\begin{figure}
\plotone{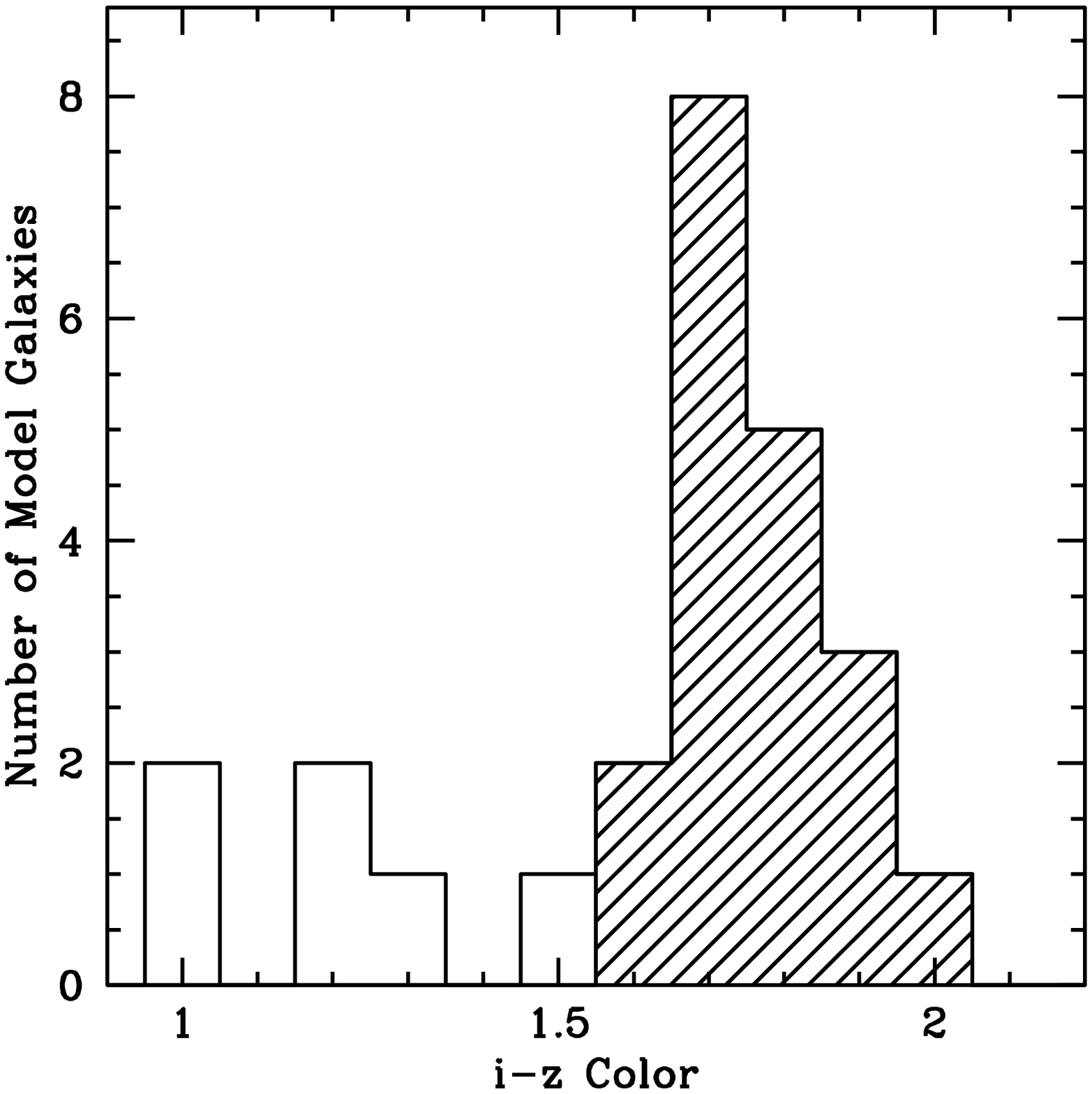}
\end{figure}

\begin{figure}
\plotfiddle{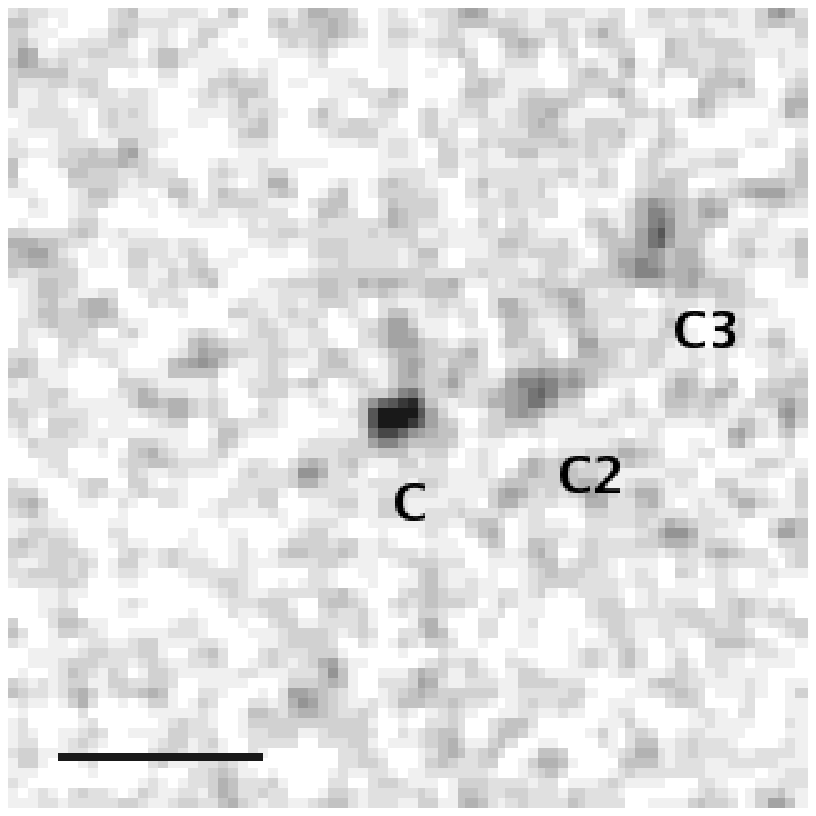}{0 in}{0}{200}{200}{-325}{1000}
\end{figure}
\end{document}